\documentclass[a4paper]{jpconf}
\usepackage{graphicx}
\usepackage{iopams}
\begin{document}
\title{Long-range interactions and information transfer in spin chains}

\author{Rebecca Ronke$^{1*}$, Tim Spiller$^{2\dag}$, Irene D'Amico$^{1\ddag}$}
\address{$^1$ Department of Physics, University of York, York, YO10 5DD, UK}
\address{$^2$ School of Physics and Astronomy, E. C. Stoner Building, University of Leeds, Leeds, LS2 9JT, UK}
\ead{$^{*}$rr538@york.ac.uk, $^{\ddag}$t.p.spiller@leeds.ac.uk, $^{\ddag}$irene.damico@york.ac.uk}

\begin{abstract}
One of the main proposed tools to transfer information in a quantum computational context are spin chains. While spin chains have shown to be convenient and reliable, it has to be expected that, as with any implementation of a physical system, they will be subject to various errors and perturbative factors. In this work we consider the transfer of entangled as well as unentangled states to investigate the effects of various errors, paying particular attention to unwanted long-range interactions.
\end{abstract}

\section{Introduction}

The ability to reliably transfer quantum information in a solid state system is one of the key ingredients to quantum information processing, particularly when considering the construction of networked and distributed systems on a larger scale. Spin chains are able to respond to this challenge, while the mathematical framework that they are based on can be applied to a variety of physical systems. Examples of successful applications include for instance encoding into soliton-like packets of excitations \cite{osborne2004}, electrons or excitons trapped in nanostructures \cite{damico2007,damico2006,niko2004}, strings of fullerenes \cite{twamley2003} or nanometer scale magnetic particles \cite{tejada2001}.

Previous work on linear spin chains has shown that in order to ensure perfect state transfer (PST), the coupling strength $J_{i,i+1}$ between two neighboring sites $i$ and $i+1$ on a chain of length $N$ should be pre-engineered according to \cite{chris2005}
\begin{equation}
	J_{i,i+1}=J_{0}\sqrt{i(N-i)},\label{PST}
\end{equation}
with $J_{0}$ the characteristic coupling constant. The time-independent perfect-transfer Hamiltonian, where we assume that any single excitation energies are site-independent, is thus:
\begin{equation}
\label{hami}
{\cal{H}} = \sum_{i=1}^{N-1} J_{i,i+1}[ |1\rangle \langle 0|_{i} \otimes |0\rangle \langle 1|_{i+1} + |0\rangle \langle 1|_{i} \otimes |1\rangle \langle 0|_{i+1}].
\end{equation}
This Hamiltonian also preserves the number of excitations. In order to assess the quality of the state transfer of entangled states, we use the entanglement of formation (EoF) \cite{wootters1998}. For unentangled states, we define a fidelity $F$ corresponding to mapping the initial state $|\psi_{in}\rangle$ over a time $t$ into the desired state $|\psi_{fin}\rangle$ via the chain natural dynamics
\begin{equation}
	F=|\langle \psi_{fin} |e^{-i{\cal{H}}t/\hbar}| \psi_{ini} \rangle|^{2}, 
\end{equation}
so that PST is achieved when $F=1$. This system dynamics leads in particular also to the so-called mirroring rule \cite{albanese2004}, which guarantees perfect transfer of any one state to its mirrored image with respect to the centre of the chain. We define the time it takes for a state to then return to its original form as the system or revival time, $t_{S}$.

\section{Effect of perturbations}

When considering perturbations, we consider fabrication defects in the spin chains themselves on the one hand, and on the other also the effect of non-synchronous or imperfect input operations \cite{me1}. For all perturbations, we consider three types of initial input states: (i) unentangled states (such as $|\psi_{in}\rangle = |110000\rangle$), (ii) initially entangled states (such as $|\psi_{in}\rangle = 1/\sqrt{2}(|100000\rangle+|010000\rangle)$) and (iii) states where the entanglement is created by the system dynamics (such as $|\psi_{in}\rangle = 1/2(|000000\rangle+|100000\rangle+|000001\rangle+|100001\rangle)$). After reviewing the effect of various perturbative factors, we will focus on the central topic of this contribution, the effect of long-range interactions.\\

\subsection{Imperfect input operations}

In the vast majority of quantum information protocols, it is assumed that the system under investigation is governed by a universal, perfect clock. In an experimental context however, this might be overly optimistic, especially when the preparation of an input state involves accessing two or more separate sites of the chain simultaneously. For a type (i) chain, this is relevant when there are multiple excitations involved as otherwise the entire system dynamics is merely shifted in time. The actual effect of non-synchronous input depends greatly on the input mechanism used. We consider the two main possibilities for injection into spin chains, which are injection via SWAP operation (an additional particle is injected into the chain, e.g. via a waveguide) and excitation via a Rabi-flopping pulse (e.g. flipping the spin of an electron confined in the chain) \cite{me1}. Both these methods allow for some correction via measurement of the system but we find that with increasing delay times between injections, systems with Rabi flopping perform noticeably worse than those where the injection was done via SWAP operation. An exception are input states of type (iii), where the access sites are far apart and there is therefore virtually no difference between the effects of the two injection mechanisms.

\subsection{Random noise}

It has to be expected that any fabrication process of spin chains may lead to random, but time-independent errors in the coupling values of the system. We model these by adding to all non-zero entries of the Hamiltonian a random energy $\eta d_{l,m}J_{0}$ for $1 \le l$,$m \le number\,of\,basis\,states$. The scale is fixed by $\eta$ which we set to 0.1 and for each $l \le m$ the different random number $d_{l,m}$ is generated with a flat distribution between zero and unity. In order to preserve hermiticity, we set $d_{l,m}=d_{m,l}$. The specific weight of the noise would have to be determined depending on the experiment under consideration, but we found that for a perturbation of 10\% of $J_0$, near-perfect transfer is still achieved during the first few periods of the system for all considered types of initial input.

\subsection{Single excitation energies}

We previously assumed that single excitation energies are site-independent, and thus did not need to be explicitly included into the Hamiltonian (Eq. (\ref{hami})). However, local magnetic fields or additional single site fabrication defects may result in a loss of this independence, and so we would have to add to (\ref{hami}) the term
\begin{equation}
	H_{1} = \sum_{i=1}^{N} \epsilon_{i} |1\rangle \langle1|_{i},
 \label{enerc}
\end{equation}
where $\epsilon_{i}$ is not independent of the site \textit{i} any more. We find that this perturbation has little effect on all input types for small values of $\epsilon_{i}$, resulting in a loss of less than 5\% for $\epsilon_{i}\leq0.1 J_{max}$, $J_{max}$ the maximum value of the coupling between neighbouring sites, at the first revival. However for larger values the effect becomes very detrimental, in particular for type (iii) input.

\subsection{Excitation-excitation interactions}

Spin chains containing multiple excitations may also be prone to interactions between excitations in neighbouring sites. To represent this, we consider the perturbative term
\begin{equation}
 \label{interc}
	H_{2} = \sum_{i=1}^{N-1} \gamma J_{0} |1\rangle \langle1|_{i} \otimes |1\rangle \langle1|_{i+1}.
\end{equation}
An example of this would be biexcitonic interaction in quantum dot-based chains \cite{damico2001,rinaldis2002}. While this perturbation is not relevant for input type (ii) (due to there only being a single excitation), we find that for the other two types, the transfer fidelity is again very well preserved for values of $\gamma$ up to $0.1 J_{max}$, resulting in less than a 5\% loss of the original input state at revival time. It can also be shown that the loss in fidelity is exponential in $N$ with Gaussian dependence on the characteristic noise parameters \cite{me1}.

\subsection{Next-nearest neighbor interactions}

In our original Hamiltonian (\ref{hami}), we assumed that interactions would be restricted to the nearest neighbour of any one spin, which is a fair assumption if the transfer between sites is based on tunnelling. If however instead we consider dipole-dipole interactions, we can model the resulting perturbation as
\begin{equation}
\label{hami2}
\nonumber H_{3}= \sum_{i=1}^{N-2} J_{i,i+2}[ |1\rangle \langle 0|_{i} \otimes |0\rangle \langle 1|_{i+2} + |0\rangle \langle 1|_{i} \otimes |1\rangle \langle 0|_{i+2}],
\end{equation}
with $J_{i,i+2}=\Delta(J_{i,i+1}+J_{i+1,i+2})/2$. We have confirmed in \cite{me1} that this is a realistic model for example for semiconductor quantum dots with excitonic qubits. Of all the perturbations considered so far, this is the most influential. Again, the loss in fidelity is exponential in $N$ with Gaussian dependence on the characteristic noise parameters. Not only do values of $\Delta$ as small as 0.05 lead to a considerable loss in transfer fidelity (for a 10 spin chain, 40\% loss with input (i), 20\% loss with input (ii), 5\% loss with input (iii)), but even for the most stable input (iii) the periodicity of the system is entirely lost after a single period (for any chain length). As some protocols rely on continual periodicity, this is a potentially very serious issue (see also \cite{avellino2006}).

\subsection{Longer range interactions}

\begin{figure}[h]
\begin{minipage}{18pc}
\includegraphics[width=18pc]{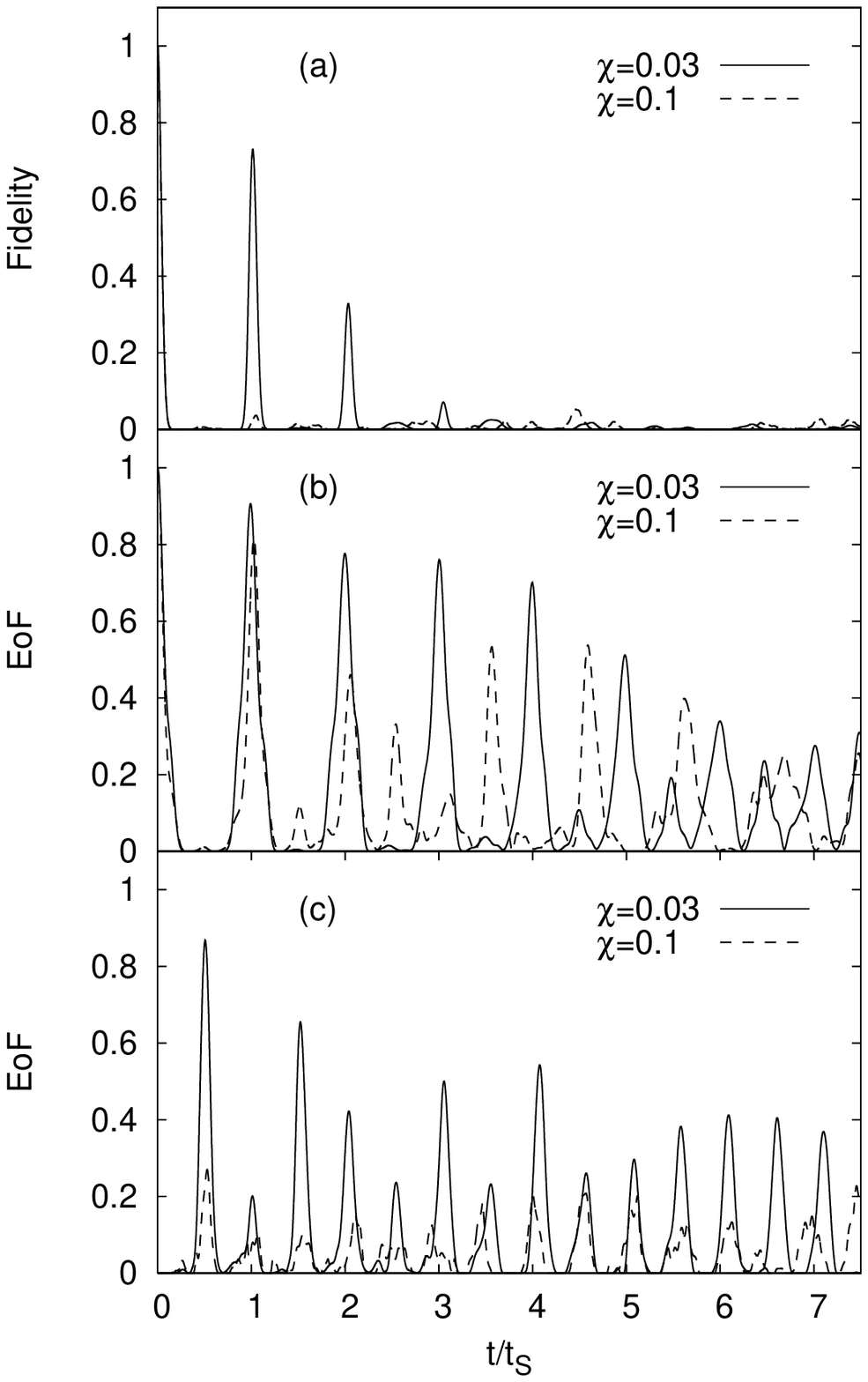}
\caption{\label{fig:i}Plots of fidelity and EoF for a 10-spin chain for two values of $\chi$ vs. rescaled time $t/t_{S}$: (a) with input type (i), (b) with input type (ii), (c) with input type (iii).}
\end{minipage}\hspace{2pc}
\begin{minipage}{18pc}
\includegraphics[width=18pc]{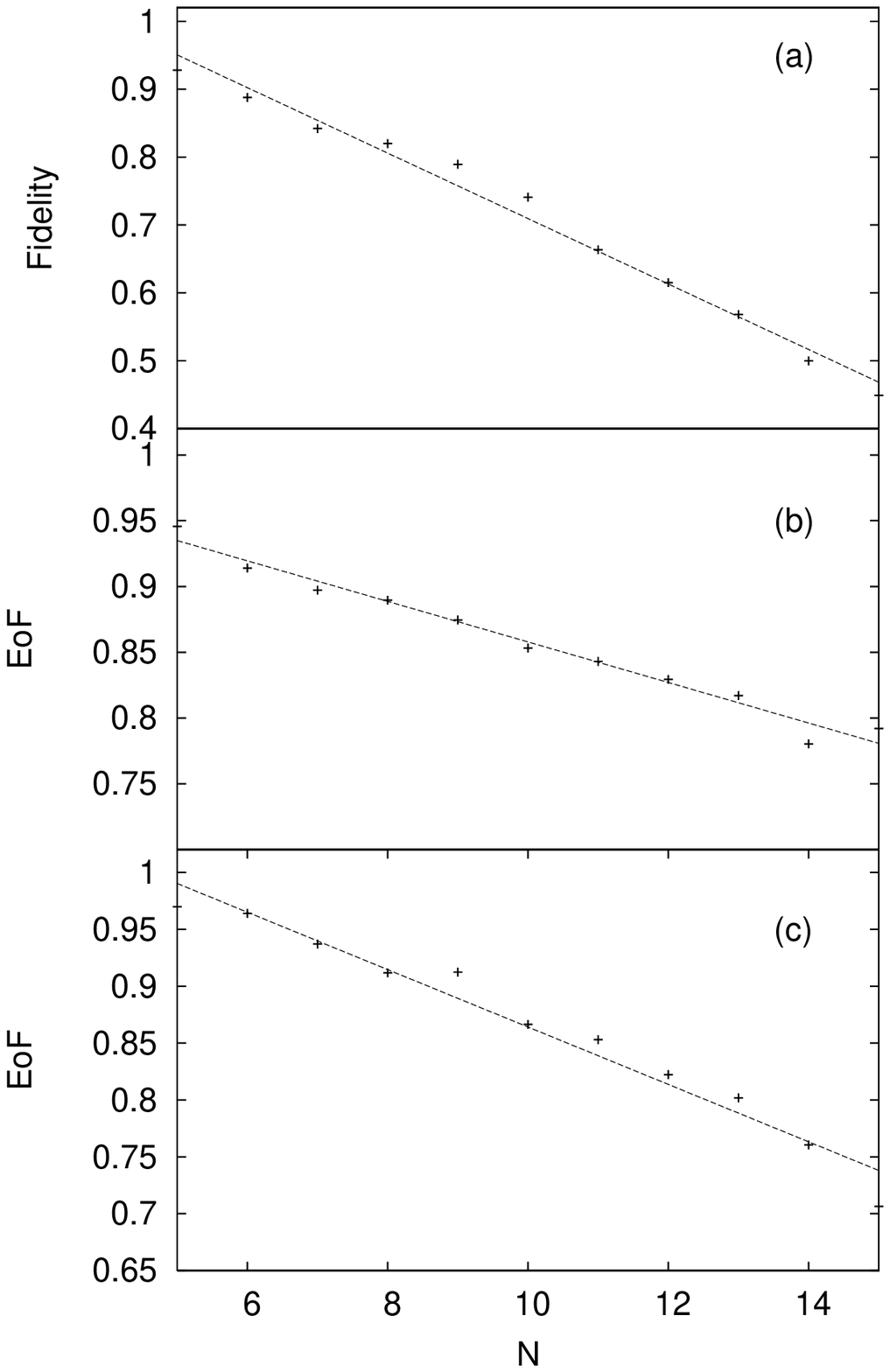}
\caption{\label{fig:ii}Plots of fidelity and EoF at the first expected revival time for $\chi=0.03$ vs N: (a) with input type (i), (b) with input type (ii), (c) with input type (iii). Lines are best fit to numerical data.}
\end{minipage}
\end{figure}

The main reason for next-nearest neighbours being the most detrimental perturbation is that it is the only perturbation we have considered so far that effectively opens up ``new channels'' for the system dynamics. We can consider this phenomenon in a more general context by formulating a perturbation factor that opens up all available channels of the system. Assuming that these unwanted longer range interactions are random, we add to all zero entries in the Hamiltonian a coupling $\chi d_{l,m} J_{max}$ for $1 \le l$,$m \le number\,of\,basis\,states$. For each $l \le m$ the different random number $d_{l,m}$ is generated with a flat distribution between zero and unity and, in order to preserve hermiticity, we set $d_{l,m}=d_{m,l}$. In figure \ref{fig:ii}, every point on the graph corresponds to an average obtained from 100 realisations.
The result of these additional channels has a different effect on the three types of initial input that we consider. In figure \ref{fig:i} we consider the dynamics of a typical realisation. We see in figure \ref{fig:i} (a) that the unentangled input type (i) suffers by far the most: even for small $\chi$ ($\chi=0.03$) the initial state fidelity is not recovered well after a single period, and subsequently entirely lost. For $\chi=0.1$, the initial state is not even recovered at all. By contrast, input type (ii) in figure \ref{fig:i} (b) conserves an acceptable amount of EoF for $\chi=0.03$ for a few periods before becoming erratic, whereas the plot for $\chi=0.1$ shows that the EoF peaks are shifted from their expected times, making the system entirely unpredictable after the first revival. A similar phenomenon can be observed in figure \ref{fig:i} (c) for $\chi=0.03$ where the EoF peaks of input type (iii) should occur at $0.5 t_{S}, 1.5 t_{S},\cdots$ but are actually starting to be shifted after just the first revival, with additional peaks forming at integer multiples of $t_{S}$. However for $\chi=0.1$, the EoF never reaches decent values. Overall it is worth noting that this is in-keeping with the detrimental effects of next-nearest neighbour interaction we have observed, where the most detrimental effect was also seen for input type (i). However the opening up of new channels even just on a small scale has almost immediate and serious consequences for all systems considered.\\
The qualitative effect of opening up new, undesired channels in a system is also dependent on the number of spins in the chain, $N$, as a higher dimensional Hamiltonian also allows for more perturbative elements. We see this confirmed in figure \ref{fig:ii}, where all three systems show a clear decreasing trend in transfer quality with increasing $N$. Again, this data was sampled and averaged over 100 random runs, keeping a constant $\chi=0.03$. We see in figure \ref{fig:ii} (a) that again, it is the unentangled input type (i) that suffers most from this perturbation, showing that, for $N=15$, the first revival is barely over 40\% of the initial input state. Input types (ii) and (iii) perform better, with type (ii) in figure \ref{fig:ii} (b) losing less than 25\% EoF for $N=15$ and similarly with type (iii) in figure \ref{fig:ii} (c) losing less than 30\% EoF for $N=15$. We see therefore that unentangled states are much more prone to the detrimental effects of new open channels in a system, but also that shorter chains always perform much better than longer ones, regardless of the initial input.

\section{Conclusions}

We have considered the performances of the transfer of both unentangled as well as entangled states under various physically relevant perturbation factors. Overall, we found spin chains to be a very robust system, provided that the errors we considered remain of the order of a few percent of the characteristic coupling constant and that imperfect input will be corrected via subsequent measurement of the system. In particular, we were concerned with the opening up of new channels for the system dynamics, beyond next-nearest neighbour interaction, as this had proved to have the most detrimental effect on PST of any type of input. We have shown that longer chains suffer significantly more from this type of perturbation, with unentangled initial states being particularly affected. Large errors lead to a complete loss of the periodicity of the system, while small errors are tolerable up to a minimum of the first revival for all input types considered. Our studies demonstrate the impact of various potential difficulties with the implementation of spin chains, while giving encouraging results for modest errors. This demonstrates that spin chains are an encouraging candidate for future experiments on information transfer in solid state quantum information systems.\\
\\
RR was supported by EPSRC-GB and HP. IDA acknowledges partial support by HP. IDA and RR acknowledge the kind hospitality of the HP Research Labs Bristol.
\section*{References}
\bibliography{papers}

\end{document}